Rethinking the Experiment

Mihai Nadin

**Abstract**: The crisis in the reproducibility of experiments invites a re-evaluation of methods of inquiry and validation procedures. The text challenges current assumptions of knowledge acquisition and introduces G-complexity for defining decidable vs. non-decidable knowledge domains. A "second Cartesian revolution," informed by and in awareness of anticipatory processes, should result in scientific methods that transcend determinism and reductionism. Physics and physics-based disciplines convincingly ascertained themselves by adequately describing the non-living. A complementary perspective should account for the specific causality characteristic of life by integrating past, present, and future. Knowledge about anticipatory processes facilitates attainment of this goal. Society cannot afford the dead-end street of reductionism. Science, itself an expression of anticipatory activity, makes possible in our days alternative understandings of reality and its dynamics.

**Keywords**: knowledge, experiment, reproducibility, decidability, non-deterministic, anticipation



To save science [1], when it is needed more than ever, requires the re-examination of some

fundamental assumptions informing scientific activity. Failed reproducibility, the crisis of this

hour, such as in biomedical sciences [2] [3, 4], or, notoriously, in psychology [5], affects more

than the validation of experiments. As a matter of fact, the experimental method in its standard

formulation becomes questionable. For science, social, economic, and political consequences

notwithstanding, to continue on the same path without questioning the premises that lead to the

current breakdown, is prone to self-destruction [6]. It is true that "Science has been peculiarly

resistant to self-examination" [7], but "metric incentives", i.e., quantifying goals and rewards,

will not change the situation. They would further instrumentalize a questionable perspective.



The situation in which science finds itself is comparable to the one that corresponds to the "flat Earth" view, which even after Ptolemy retained some currency.

*Data about the reproducibility crisis*

**Thesis 1**: Experiments intended to advance knowledge of the living are useful but not reproducible.

Against the background of successful technological innovation grounded in physical science, the life sciences, while seeking legitimacy in the guise of chimeric experimental replication, are delivering below expectations and societal need. If the object of the experiment is the physical substratum, reproducibility can be expected, provided that the experiment is properly designed and carried out. For instance, one could re-wire a genome (introduce new links between unrelated genes), or measure an interaction between biological components with a high-degree of precision, similar to how a chemist would study the attractive forces between non-living elements. But such experiments contribute to knowledge of physics or chemistry, not to the life sciences, whose object is change in the living. Moreover, the same is not meaningful for exploring, for example, protein folding or anticipatory genetic expression, or for that matter evolutionary dynamics. We shall examine why this is the case.

 In contradistinction to experiments in "physics or astronomy or geology" [8] [9] — knowledge domains identified as test provable by the vast majority (90%) of researchers —failed reproducibility occurs almost exclusively in life science experiments. From the many reports published, we learn that in particular domains, 80% of published results from researchers who earned the respect of their peers proved to be irreproducible. In one review [10], findings from the biotech company Amgen are detailed. Up to 100 of its researchers attempted to corroborate



data from 53 cancer research reports from well-known facilities, published in leading journals. Only 6 of the studies proved to be reproducible (i.e., ca. 10%). The pharmaceutical branch of the Bayer conglomerate had no better luck in seeking validation for research in oncology, women's health, and cardiovascular medicine.

The crisis of reproducibility has undermined the credibility of leading scientific publications (*Nature*, *Science*, eLife, PNAS, etc.)—none eager to address fundamental science. They literally censor contributions to the subject not in line with the views they promote. Screening by staff eliminates the opportunity for peer review. For all practical purposes such publications have become self-styled newsletters for the extremely profitable industry of experiments. They preach to the choir instead of offering a platform for scientific debate. One example: lizard mobility framed within the momentum conservation principle of physics [11]. A lizard-like robot looks good on a journal cover. It does not matter that the experiment does not address the real subject. The Radio Shack toy conveniently made into a "lizard" robot proves the physics. A 2-dimensional tail modeling confirms (through circular thinking) the false hypothesis. Anticipatory aspects of the lizard's mobility were fully ignored. A review of this experiment, submitted to Nature, never made it past the screening.

The crisis undermined as well the activity of funding agencies (governmental or private). Some, such as the national Institutes of Health (NIH) rushed to issue new grant guidelines without understanding where the problem lies. Guided by a concept of knowledge acquisition generalized from physical science to the living, they disburse public money for more research equipment, but not for appropriate hypotheses. Stimulating alternative ways of thinking seems to no longer be part of their mission. Talent is wasted on servicing expensive machines for data acquisition instead of advancing new ways of thinking. And when the cheap labor of graduate



students is not available, the jobs are outsourced to scientists in countries desperate to have access to new machines. I experienced this in several laboratories, but most painfully at the Bechtereva Institute of the Russian Academy in St. Petersburg, where providing service in data acquisition replaced the once respected original work on advancing hypotheses regarding the brain.

The American National Academies, as well as the British Academy of Medical Sciences, the Max Planck Institute and many other prestigious institutions, were prompted to examine the disturbing facts [12]. A new field of inquiry, focused on experiment replication and reliability before the experiments are carried out is ascertaining itself [13] [14]. However, we know that "when a measure of success," such as metrics "becomes a target, it loses its significance" [15]. Be this as it may, this crisis should not go to waste into more pseudo-science (based on the ever fashionable probability theory, Bayes, or some fancy mathematics) about bad science, or how to stifle science by further institutionalizing rules and regulations soon to become a goal in themselves. Without addressing the origins of the problem, the scientific community will only continue to reproduce the never proven assumptions upon which the majority of its activity is still carried out.

*Addressing a systemic condition*

The replication quandary is an opportunity not to be missed. The relation between various knowledge domains and the need to adapt research methods to the specific dynamics of the subject that scientists attempt to describe is an unavoidable subject. Opening an in-depth discussion of this subject, as we shall try herein, would be in many ways a promising beginning.



Data dredging, omission of null results (nobody wants to hear about them), underpowered studies, and underspecified methods or weak experimental design are symptomatic of weaknesses brought up in the discussion of replication failure. A small-size sample saves effort but undermines robustness. Open data, always desirable, more collaboration, automation—where human error can be avoided without altering the meaning of the outcome—and similar methodical suggestions ought to be considered. But ultimately they are not the answer. Therefore, now more than ever before the scientific community has to come to a rather sobering realization: eliminating weaknesses such as those mentioned will not change the systemic condition that resulted in failed replication in life sciences. The lack of reproducibility is only a symptom of a deeper reaching malaise: the refusal to accept alternative conceptions of reality and its extremely rich dynamic forms.

As opposed to the non-living, the living is endowed with control processes manifest at each of its levels—from cells to organism to interactions with the world. There is freedom at each level, and there are interactions expressed as constraints. What I describe here goes beyond the hierarchy theory (of Pólya, continued by Pattee and Rosen, among others). My preference is the Principle of Minimal Interaction [16]: interaction among constituents at a lower level of the organism hierarchy follows the path of minimizing external input. For instance, in motoric expression, each joint is under its own neural control. Local interaction among elements is such that the outcome (motoric expression) is minimally dependent on the output of other elements. On the global level, the outcome of the structural unit (e.g., elbow joint) is minimized by changes in the output of other elements (the other joints in the kinematic chain).

To the best of my knowledge nobody involved in the evaluation of the situation (by no means new) has issued a call to the scientific community to re-evaluate the underlying assumptions



upon whose basis knowledge acquisition and confirmation are pursued. The scientific community ought to come to the realization that experiments different from those that undergird the progress of physics, chemistry, astronomy, geology and the like are *unavoidably* non-reproducible. Science practitioners have tacitly accepted reproducibility since the early stages of the Cartesian grounding of the experimental method, i.e., the reductionist-deterministic model. Experiment was always congenial to inquiry; reproducibility affirmed an expectation that became the epistemological premise: determinism. This was never a matter of philosophy, as some frame it, but one of practical consequences. Replication of experiment, or for that matter of medical treatment, has become a matter of public concern because it is not about one or another scientist, or physician, missing the expected threshold of acceptance. This is about failed science in an age of higher than ever expectations, given the significance of knowledge in general, and, in particular, of the living, for the future of humankind. The critical re-evaluation of the epistemological premise is the only rational path left to pursue.

Indeed, machines can be built (and were successfully built) on account of physics and chemistry. They are supposed to be as pre-determined as possible. Their functioning is repetitive. But during the timeline of the machine revolution, the understanding of life has improved only slightly. Everything that lends itself to the building of yet another machine can be reproduced--that's what machines are for. But they are not science about the living. That is, they are not even science about the living beings who made the machines. The lever is an extension of the arm, but not a theory of motoric expression, even less of muscles, tendons, bones, joints, etc. The artificial neuron inspired by the living neuron is a mathematical construct of extreme application potential. But it is not knowledge about the neuron that inspired it, and it is not about any neuronal processes. Albeit, we are still in a rudimentary phase of scientific development regarding aspects



of life (such as intuition, emotion, creativity, spectrum diseases, etc.) associated with motoric expression or neuronal activity (to mention only two aspects). At the same time, we benefit from the sophisticated machinery in production facilities or deployed in deep learning, inspired by the living neuron. One such machine beat the world champion in *Go* (a game more complicated than chess, but still of permutation choices in a finite space); another imitates the art of the masters of painting.

Research, instead of speculation, is a shared choice that scientists made—giving science its impetus. Nevertheless, the expectation that research is best validated through reproducible experiments, no matter what the subject or purpose, became questionable. The empirical evidence accumulated suggests the need to re-examine this expectation. In view of progress in science, it is only logical to think that reductionist and deterministic explanations are begging for complementary perspectives. This in itself is an argument that cannot be ignored for the suggested re-examination. The understanding of what Newton called *Nature*, under which label he aggregated both the physical and the living, might prove as inadequate in our time as it was when it was articulated.

*A question discarded*

After vitalism was debunked, and replaced by "mechanism" [17], science rejected the distinction between the living and the non-living. This rejection is quite surprising, since in science you don't discard a question because it was improperly answered, or because the alternatives answers are not aligned with a dominant view. The foundational works of Walter Elsasser [18], and Robert Rosen [19], not to mention Erwin Schrödinger [20], advanced views of nature different from those of Descartes and Newton and their followers. Their contributions



were pretty much ignored at the time they were published. Elsasser and Rosen articulated arguments that were quite different in their perspectives: Elsasser within a physics-inspired perspective, Rosen in mathematical language. Both deserve a closer look at this moment of questioning the research and validation methods of life sciences. Both provided proof that the living is heterogeneous, purposeful, and anticipatory, as opposed to the non-living, which is homogenous, purpose-free, and reactive. These distinctions have certainly earned the attention of scientists—even those who do not read any reports older than five years. If, indeed, to know is to be aware of distinctions—especially those of a fundamental nature, corresponding to their different dynamics—such distinctions cannot be eliminated by fiat—or worse, just ignored.

While physics and physics-based theories adequately describe the non-living, there remains a need for a complementary perspective that expresses the nature of life. What defines this perspective is the fact that the specific causality characteristic of life is accounted for by integrating past, present, *and* possible future. The living changes in a way different from the non-living. The causality characteristic of the living is much richer than that of the physical, if for no other reason at least because the living is reactive (like the physical) and anticipatory (which the non-living is not). However, the scientific community, conditioned by an education set on the Cartesian foundation at the expense of any alternatives, is reluctant to accept this. Moreover, the description of change in the living calls for particular means and methods to properly capture it. Measurement and, by extension, the limited model of experiment, appropriate for describing physical non-living entities return incomplete, and at times confusing, knowledge when applied to capturing life change.

Causality, i.e., how and why things change regardless of their specific nature, proved to be richer than what classical determinism ascertained. Under "things" belongs not only a stone that



eventually turns into sand, but also earthquakes, the sequence of seasons, the day-and-night cycle, women's monthly cycle, the way humans think, the intelligence of plants, the adaptive nature of the microbiota (to name only a few). The matter from which physical entities (not endowed with life) are made remains the same, subject to what the laws of thermodynamics describe, in particular increased entropy. The living, from the simplest monocell to the human being is in an uninterrupted state of remaking itself, *sui generis* re-creation of its constitutive cells. It is neg-entropic. The re-making, i.e., renewal, of cells takes place at various rhythms: some are renewed almost daily; others over weeks, months, years, and others not at all.

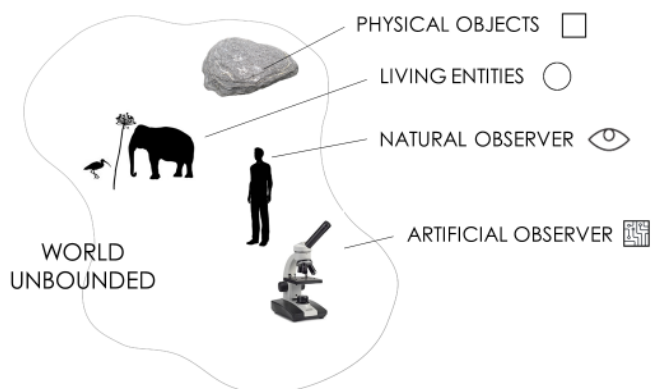

Fig. 1. Unbounded world as open system

Determinism, the characteristic causality of physical phenomena, is convincingly relevant to the physics and chemistry of the living. Its description, for instance, through experimental data, returns an incomplete explanation of the specific nature of the changing living. Just to present an example along this line: physical forces (e.g., pulls, compressions and stretching, distortions) applied to a cell can further affect it, probably more than the inherited genetic code does [21, 22]. Taking both physical forces and the genetic code into consideration affords an understanding of



cell changes that neither can deliver alone. Non-determinism, describing a relation between cause and effect that takes the form of a multitude of possible outcomes, sometimes contradictory, pertains to change as an expression of something being alive, influencing its own change. Changes due to physical forces applied on cells (think about cutting yourself with a sharp knife, or falling against a rock) and genetic processes governing dynamics are interwoven. There is no way to unequivocally predict whether a cell becomes cancerous or simply divides in a process of self-healing. This is neither randomness nor stochastic expression—place holders for the notion of non-determinism, which is almost never acknowledged in its full expression (different from randomness).

To know how the physical changes is to infer from a quantitatively described past state to a future state, under assumptions usually defined as initial conditions (also expressed numerically). Knowledge of the process underlies our ability to predict a future state. One description, sometimes rich in detail, captures the process. To know how the living changes requires more than the physics-based description or what chemistry, usually associated with life, ascertains. The empirical observation of changes in the living always leads to multiple descriptions, corresponding to multiple possible states, sometimes simultaneous, none exclusive of the other. An adequate explanation of change in the living requires integration of inferences from past states with interpretations of the meaning of possible future states together with the possible paths to them. Anticipation is expressed in actions informed by possible future states [23]. The domain of the possible is by no means less real than that of what is (the extant). The ontology of the living entails that of the possible. A suggestive analogy is justified here: potential energy is no less energy than that at work in a physical or living process.



The framing of change within the respective consequences, different in the physical and the living, is key to understanding their difference. The causality specific to interactions in the physical realm, transcending Galileo's world, is described in Newtonian laws—action-reaction, in particular. It was further refined in relativity theory, and for the micro-level of matter in quantum mechanics.  The causality specific to interactions in the living, which is purposeful, includes, in addition to what the laws of physics describe quantitatively, the realization of significance—What does it mean?—in connection to the possible future. For instance, the question "What does it mean?" is the implicit question addressed in each synapse. Photosynthesis is the outcome of meaningful processes, not of data processing at the molecular level. Numbers do not play any role; they do not even exist at that level. We can use them to describe the process (or at least part of it). At the neuronal level much guessing takes place, sometimes felicitous, sometimes not. The same holds true for cell interaction. Protein expression allows for electron transfer across membranes, but this does not make the process machine-like (as Dutton [24] is inclined to describe it), and does not make electron transfer a characteristic of life. The construct called *number* (to which we shall return) emerges at a higher level than neural activity or cellular expression. The construct called "machine" is the outcome of mathematics, and not a characteristic of life. As mentioned above, we can use numbers to describe them, but in doing so, we need to realize that such descriptions (i.e., representations) are incomplete. Of extreme significance is the fact that the living, in addition to the constraints of physics, or those associated with molecular chemistry, is subject to contingent rules of behavior, expressed as anticipatory action. This is usually brushed aside, or trivialized (as in identifying historicity and contingency [25].



*How did we get here? The theology of reductionism*

The consensus view was summed up in the Preface to a *Committee Report to the National Academy of Sciences* in 1970: "Life can be understood in terms of the laws that govern and the phenomena that characterize the inanimate, physical universe, and indeed, at its essence, life can be understood only in the language of chemistry." This became the official doctrine of how science would be carried out in the USA, pretty much like dialectical materialism was in the Soviet Union. Those who still questioned the reductionist position—they were not fired or jailed for this—received an encouraging face-saving line: "Until the laws of physics and chemistry had been elucidated, it was not possible even to formulate the important, penetrating questions concerning the nature of life." Philip Handler [26], at that time president of the National Academy of Sciences, published these thoughts in the "Preface to Biology and the Future of Man—a National Research Council Report." Thus, what used to be the dominant but not exclusive view among scientists became the dogma of science (like the "flat Earth" was until Ptolemy, and even past him). The 1989 Opportunities in Biology report [27] obviously reflects a new, "state-of-the-art" (mentioning "recombinant DNA, scanning tunneling microscopes, and more") focus on the molecular scale of life. Nevertheless, it remains aligned with the position adopted almost 20 years back. It should not surprise that the current state of the art in knowledge acquisition and dissemination pertinent to the living—another 30 years later—is evidently the consequence of the official "theology" of science. It became a form of science fundamentalism. Of course, such a drastic characterization, which has to be extended to the ever-deepening machine reductionism, begs to be well grounded.

The machine model of reductionist determinism, made explicit in the Cartesian perspective, was the clock, followed by hydraulic, pneumatic, electric, and all kinds of engines. In our days,



they were replaced, in their role as the model of the human, by the computer (the 1989 report takes note of it), and are expressed as algorithmic reductionism. It is quite useful to understand the process. Machines originate in the making of tools. Starting with the lever, the pulley, the inclined plane, machines were made, like tools before, as extensions of the body itself. They were supposed to ease labor and make it more efficient. All other machines came into existence as constructs of the same nature: imitate some human activity with the aim of augmenting the output of labor; or, as with the computer, imitate the brain and its output. The making of something that seems to acquire qualities beyond those endowed upon it is a subject of *dialectic*—a reasoning method for establishing the truth of arguments. The Young Hegelians of the 19th century extended this discussion to the dialectics of ideas. They were aware of the fact that words, or numbers, or languages (including those of science) could take on the appearance of being independent of those who originated them. In discussing the Christian god, they also took note of the dialectic of making tools and relating to them: tools seemed more powerful than their makers, as though they were endowed with what appeared as superhuman qualities. The same dialectic of self-deception (i.e., they seemed more powerful than their makers, as though they had magical properties) was at work in creating social institutions and political systems. An inversion takes place: "God" was made by humans in order to explain what appeared as transcending the individual, or even the group sharing in the naming. Such a construct is useful for explaining phenomena of a scale different from those involving direct interaction or human reasoning. After constructing the entity "god," "God" appears to be the originator of those who made it (or wrote commandments attributed to it), as though it existed independent of the maker, moreover as the Maker of its makers.



The machine was meant to augment human abilities. It is an expression of applied knowledge, informed by human creativity. In performing better than the humans who made them, and even independent of them (the hydraulic machine, for instance, or the pendulum, or the computer), machines appeared as though they were the blueprint of life itself. The understanding of the heart, the lungs, or the brain as machines led to the inverse statement: the heart (or brain, or lungs) is nothing but a machine. With the advent of the "machine of machines"—the computer—the human being became the tool of its own tool—a reduction. This understanding is objectified in ideology—the logic of our ideas—pretty much like that declared in the theology of physics and chemistry, to which the living has been reduced. Like the god concept, it was never proven that the reduction to physics and chemistry—given that the living is embodied in matter—or the reduction to machines, affords a better understanding of what life is, or that it represents more than analogy. The belief in a god results in self-reinforcing ideas and actions: for instance, living according to commandments, or getting married, or finding peace with oneself. The recently proclaimed "miracle medicine," extending life by 20-30 years [28] [29] exemplifies the idea (based on the just released account issued by the Harvard School of Medicine—even though the same has been known for a long time). The same process of self-reinforcement holds true for machines and physico-chemical reductionism—it's a placebo effect, which is a particular form of anticipation expression.

Concepts and views that document the continuity of questioning the premises that resulted in the current crisis of reproducibility, and implicitly in questioning a rather limited understanding of experimental evidence, make up a large body of contributions from scientists and philosophers. Let us take note of the fact that neither Galileo's mechanics, nor Newton's theory, nor Einstein's, nor the still under-defined quantum mechanics is the outcome of experiment. That



experimental evidence confirmed them is beyond question. The expressive power of deduction exceeds that of incomplete inductions embodied in experiment. Each of the descriptions mentioned corresponds to entailed processes in the world. That is the source of their predictive power: a future state can be inferred from a past state, provided that we consider initial conditions. Darwin, aiming at being the Newton of biology, based his evolution theory on ample empirical observations (his own as well as those made by others over time). Nevertheless, evolution is un-entailed [30]: one cannot pre-state a future evolutionary state, like physics pre-states the future position of a particle or of a rocket. Gould [31] dealt with the contingency of evolutionary processes. The "replaying the tape" thought experiment he proposed, in reference to experimental data from the seas of the Burgess Shale, implicitly affirmed that there are no biological laws. That is, biology is idiographic, in contrast to physics, which is nomothetic (we shall return to this). Discussions of Gould's views [32, 33], [34], [35], [36], are relevant to what we discuss because they brought up experiments. One of them is the long-term evolution project (using evolving populations of E.coli bacteria [37]); the other concerns macro-evolution [38]. Without entering into details, let us take note of the fact that these experiments showed that the closed system of the experiment makes the expression of holistic dependencies characteristic of the open system of nature impossible. They are successful for reproducing premises, but not for documenting contingency, or undermining Gould's views. "Convergent evolution as natural experiment" [39] addresses iterated evolutionary outcomes, conceding that histories can be generalized, but are not law-like generalizations.

This in itself defines an epistemological horizon: in order to cope with change, including their own, humans describe it, hoping to find in the description clues concerning possible consequences. Literature documents how descriptions evolved from the pictorial shorthand



associated with incipient humankind to language and to progressively more abstract representations [40], the languages of science.

*Knowledge is representation*

Knowledge, in its variety of forms (transitory, conjectural, implicit, explicit, instinctual, emotional, rational, etc.), is the outcome of learning. The living, in all its known forms, learns continuously. In some ways, evolution is the aggregate expression of how learning supports life in its continuous change. Knowledge acquisition is the expression of anticipatory action: the contingent, possible future explains the need for knowledge. The process is non-deterministic: it can, when the knowledge informing life actions is meaningful, reinforce life changes; or, when it is not appropriate, undermine them.

To explain how knowledge is acquired along the timeline of life is to pursue a never-ending spiral [41].

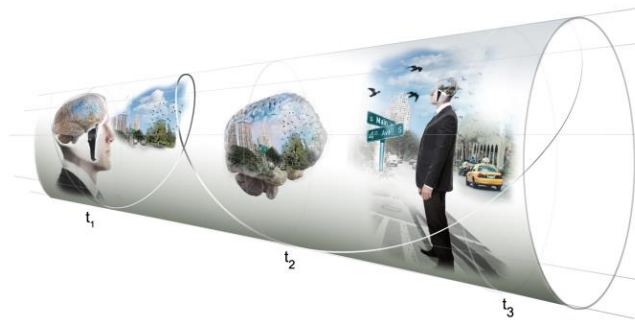

Fig. 2. Knowledge expresses the open ended learning experience of the living

Knowledge of physics or chemistry, that is, at a level of generality higher than that characteristic of cell activity, of tissues, of organisms, of individuals, is the expression of anticipatory action in regard to the living and non-living component of reality. Physico-chemical aspects of the



interaction between the living and the world (once again: living and non-living) are part of a larger dynamics, in which they are necessary but not sufficient conditions for knowledge acquisition. In this process, in addition to processing what is, the living generates a new reality, i.e., representations, upon which it acts.

Representation is not neutral. In the representation, the represented is reduced to whatever is intentionally, or accidentally, of interest, to what is significant. The illusion that a representation, such as a number, is objective, independent of the representer, is the source of the many "religions" developed inside science over time. A second illusion is that representations are like mirror images of what they represent. In reality, each representation is also constitutive of what it references. Words make reality just as tools make reality. Time representation, as interval clocked by a machine, is one of the most obvious examples. Rosen [42], fully dedicated to finding more adequate descriptions of life (such as his model theory) worked hard to prove that the origin of the measurement method corresponds to the mystical understanding of numbers in Pythagoras's idealism. Paradoxically, the idealism of the number, i.e., the idolatry of numbers, undergirds the experimental method meant to reject idealism. In the broader epistemological perspective of science, the goal is to achieve objectivity: how to know the world independent of who tries to understand and describe it, and independent of the means to represent it. The fact that the description ultimately expressed in a language (natural or artificial) makes objectivity dependent on the relation between expressive power of language and its precision has to be taken into consideration. The more precise language of Yes (1) and No (0, zero), "spoken" by digital machines, is more "objective" than the ambiguous natural language spoken by scientists expressively (one can even say *artfully*) describing natural phenomena.



In search of objectivity, the "knowing" subject had to be done away with. For physical processes, such as the falling of a stone, or for that matter a lunar or solar eclipse, it really does not matter who observes them. The equations, the sketch, the picture, the video, the animation, and the sentences describing them capture different aspects of the process. The outcome of experiments pertinent to the law of physics describing them are independent of characteristics of human perception. The experiments affirm an understanding of dynamics whose underlying time is nothing other than duration—between cause and effect. The gravity machine called "pendulum" (eventually packed into a clock) labels intervals, which correspond to diurnal time. It objectifies an understanding of time that assumes that there are no differences between the living and the non-living. Late in his life, Einstein realized the error of connecting time to the physical clock (i.e., a machine)—a position more or less consistent with the discovery of the limited speed of light (once again: independent of whether it is observed by a machine or a human being). Living time experiences, not reducible to their physics, are not clocked. The continuous renewal of cells in the organism exemplifies the idea. The renewal processes are timed: which sequence of events (physical, chemical, or informational in nature) or which configuration (of matter, energy, symbols, etc.) is significant for the maintenance of life, for reproduction, for creativity. Instead of duration independent of the observer, we have the *meaning* of time, pertinent to observation and to the observer. Among cells, as well as among neurons, interactions are slow. The speed of light, in respect to which duration is referenced, is actually of no significance at the cell level (cells are tightly packed). The experimental evolution project involved the cloning of bacteria in order to produce genetically identical populations— which would be an anomaly in nature. Of course, biological time was reduced to duration, and in effect the experiment became one in duration of adaption to a new, artificial, environment. The



bacteria experiment at Michigan State University pre-empted the anticipatory expression, reducing life processes to physico-chemical processes.

Time in the living of course reflects the physics of gravity. In addition, biological self-preservation prompts the anticipatory action of the organism (as a whole) as it avoids hurt in the process of falling. This anticipatory action takes place at a different timescale: faster than real time, i.e., faster that physical duration, as it is described and exemplified by recordings on film or digital media. (More on this in [43]). The physics is expressed in the description of the fall: always the same. The biology is expressed in the timescale within which anticipatory action—not always successful—allows for avoidance of getting hurt. We think faster than real time; distributed thinking engages the entire body, of many integrated clocks—to  use the machine name for labeling them—"ticking" at different timescales, many of them variable.

The incipient machine concept (de la Mettrie, Descartes), as the concrete embodiment of determinism and reductionism, was hypostasized in the larger context of an economic system dependent upon machines. Since Descartes' time, science became the religion of its own never-proven assertion concerning the living as a machine, and causality as determinisms. There was a historic necessity to machines, as there was historic necessity to the attempt to extend the rationality embodied in them to the living. The practicality of machine understanding of the world reflects that of the machine as a means of production. Since that time, experimental evidence has meant the reproduction of machine reproducibility, i.e. the "production" of knowledge, instead of discovery.

*The ideology of science*



Science, however, is not meant to align everyone and forever with the religion of immutable concepts. The supplicants of the machine religion (the machine fundamentalism) are captive to the circularity of religion: the machine and the machine metaphor are the outcome of human attempts to overcome their limitations, including those affecting the possibility of the knowing subjects to be able to know themselves. Those who claim today that our reality is the outcome of some larger machine (a computation, or a simulation [44]) are only bringing the solipsism religion farther than it ever was: we humans constructed science and now we search for experimental proof that it makes us, as we continue to search for answers relevant to our existence within circumstances so different from those reflected in Descartes' views. It is no longer acceptable in science to maintain that the living and the non-living (the physical) are different. Official science (Handler 1970), and the pursuant reports on the state of knowledge in biology, proclaims that they are fundamentally the same.

This is reflected in the beliefs of many practitioners. Whitesides [45], defining the new chemistry (recall the claim "life can be understood only in the language of chemistry") describes its new ambitions: What is the molecular basis of life? ("…life is an expression of molecular chemistry"); How does the brain think? ("thought" is simply interacting molecules, and hence chemistry). A "new physics theory of life" [46] ascertains that matter, under some circumstances, acquires life attributes: "You start with a random clump of atoms, and if you shed light on it for long enough, it should not be so surprising that you get a plant." Even Darwin's theory becomes a special case of a more general physical phenomenon. Tegmark [47] went so far as to consider the universe made out of mathematics. Barabasi [48] generalizes from the large networks (Internet, for example) to protein interactions, not realizing that living networks are continuously reconfigured. Berry et al. [49] find shapes in nuclear astrophysics quite similar to those of a



cellular organelle. Of course, for these authors the stacked sheets of neutron stars are more complex. (The jargon of reductionism always implies degrees of complexity, without ever defining them.) Latash [50:136] argues that "Behavior of biological systems is based on basic physical law, common across inanimate and living systems . . ." That anticipation is definitory of the living—i.e., transcends the features (homeostasis, metabolism, growth, reproduction, etc.) he mentions—does not fit within the view of the living he presents. However, Latash leaves the door open (not unlike the authors of the 1970 report) to ". . . currently unknown physical laws that are specific for living systems." Do they have to be physical?—is a question he, and many others sharing this view, does not address.

These are only examples. Examining the position adopted by many researchers, it is clear that the official science program of the National Science Council and of the Academy of Sciences became ideology. Even foundations claiming to pursue alternatives (such as the well-endowed Templeton Foundation) are ultimately practicing it. Suffice it to point out that, impervious to anticipatory processes, they embraced the so-called prospective psychology laden terminology of experiments impossible to replicate [51].

The official science ideology justifies a false premise, never proven, always assumed to be true (as God is assumed to be). It instrumentalized this premise, in science policy and the associated reward mechanisms, which forces those seeking support to align to the ideology. Not surprisingly, the machine that makes all machines—i.e., the computer—is exactly what the Church-Turing thesis ascertains: every physically realizable process can be computed by a Turing machine. A simple, deterministic recursively functioning device does it. This would make the claims of chemists and physicists nothing other than an expression in computation. The inference that living processes are Turing computations entails exactly the assumption of



Hilbert's formalist program: semantics (of mathematical statements) is reducible to syntax. The machine internalizes all referents to the world in syntactic form. There is no isomorphism of any kind between the state of a machine and that of a living open system. This fact still escapes the logic of the preachers of the machine religion. Moreover, there are alternatives to the algorithmic Turing machine—which Turing himself considered [52] [53], in particular the assumption that everything can be measured. This is disregarded in the super-god religion of computation, where computation and algorithmic are considered equivalent.

Rosen, as mentioned already, found fault with Pythagoras for this situation. To assume that mathematical truth, formal in nature, and at best the expression of cognitive performance upon a quantified representation of reality, corresponds to the dynamics of reality is conducive to a reduction of all there is to what can be mathematically represented, i.e., expressed in the language of mathematics. That a mathematics of living processes might be an alternative does not fit in the reductionist program. Gelfand [16] and Bernstein [54] (to whom we shall return) argued in favor of such a new mathematics. Probably along the same line of thinking is the inference that what does not align with the premise (i.e., is not algorithmic) has to be implicitly wrong. Therefore, non-determinism was declared a form of determinism, and open systems were qualified as canonically perturbed closed system. The opposite provides a more adequate understanding. Actually, there is a need for a comprehensive understanding of non-determinism, as well as for better descriptions of open systems, and of holism, as a characteristic of such systems.

Inspired by Niels Bohr, one can hypothesize that we would be better off opting for a model of complementarity. This, by extension, applies to the experiment as a knowledge acquisition and validation method (among others) for everything pertaining to the physics of the world, and



empirical evidence, as interval series (or history) for the living. On account of interval series, interpretations of the data they afford lead to the category called story, which Kauffman [55] [56] (among others) called to the attention of the scientific community. This particular aspect deserves more detailed examination [57], if indeed story, in a well-defined sense, should become the outcome of life sciences and inform practical activities (such as healthcare).

*A distinguishing criterion*

Gödel's concept of decidability (the logic pertinent to axiomatic systems used in arithmetic operations) can be applied in defining knowledge domains. It offers the possibility of describing the particular manner in which the physical and the living can be effectively distinguished by an observer (natural or artificial). Gödel's Incompleteness Theorems ascertain that any theory trying to describe elementary arithmetic cannot be both consistent and complete, i.e., is not decidable. The dynamics of physical reality is by its condition different from that of the dynamics of an evolutionary system (actually a system of embedded systems). The number of entities involved in physical processes, from the simplest (such as those described in Galileo's mechanics and in Newton's laws) to the more complicated (such as those captured in Einstein's physics or even in quantum mechanics) is finite. This number does not increase along the timeline of change. Living processes are open ended. The number of entities is in continuous expansion. They are endlessly re-created. That Gödel's theorems concern not only descriptions of reality (formal systems), but also reality might not be directly provable within the formalism in which they are expressed. However, they reflect the constructive nature of all human knowledge, about arithmetic as much as about anything else in the world. If mathematics and measurement are co-substantial, logic (as in Gödel's logical theorems) pertains to interpretation, and is co-substantial



with representation. Representations of the world change the world. The undecidable becomes, to use a figure of speech and not a precise qualifier, more undecidable. Moreover, Heisenberg's uncertainty principle [58] shows that extending Gödel's theorems to reality actually reflects our understanding of fundamental properties of material systems (he limited himself to quantum systems).

On account of these considerations we can proceed with an effective distinction procedure. The focus in this alternative view is not on Gödel's rigorous logical proof—which can be used in describing reality, and not only the numbers representing it—as it is on the notion of decidability, extended here from the formal domain to that of reality.

**Definition**: An object of knowledge inquiry is decidable if it can be fully and consistently described.

Indeed, physics, astronomy, geology (mentioned by Goodstein, as well as by many subsequent reports), knowledge domains where reproducibility of experiments is close to 100%, are descriptions of dynamics (how things change), i.e., representations of change that can be complete and consistent.

**Lemma**: Experiments involving decidable processes are reproducible. Such descriptions undergird predictions—the expected output of science without necessarily guaranteeing it.

Observation: The dynamics of the decidable—i.e., how entities that can be fully and consistently described change—is not necessarily a sufficient condition for making it predictive. The Poincaré problem regarding the 3-body dynamics is probably the best known example concerning this observation.

**Thesis 2**: The threshold from the decidable to the undecidable is the so-called *G-complexity* (G for Gödel, obviously [59]).



The source of undecidability is the interaction through which the living is identified as anticipatory. If the notion of complexity conjures any meaning, it cannot be reduced to numbers or to the language of mathematics, which captures only quantitative aspects of dynamics. G-complexity remedies the generality of the notion of complexity in favor of a distinction—the decidable—that can be probed. The cell is as undecidable as tissue, and moreover, as the organism it makes up.

**Thesis 3**: Change above the G-complexity threshold is undecidable.

Interactions are the concrete expressions of G-complexity. In the living, interactions continuously multiply. As the living returns to its physical condition (from senescent states to death), interactions decrease and settle in the decidable domain of physical and chemical interactions.

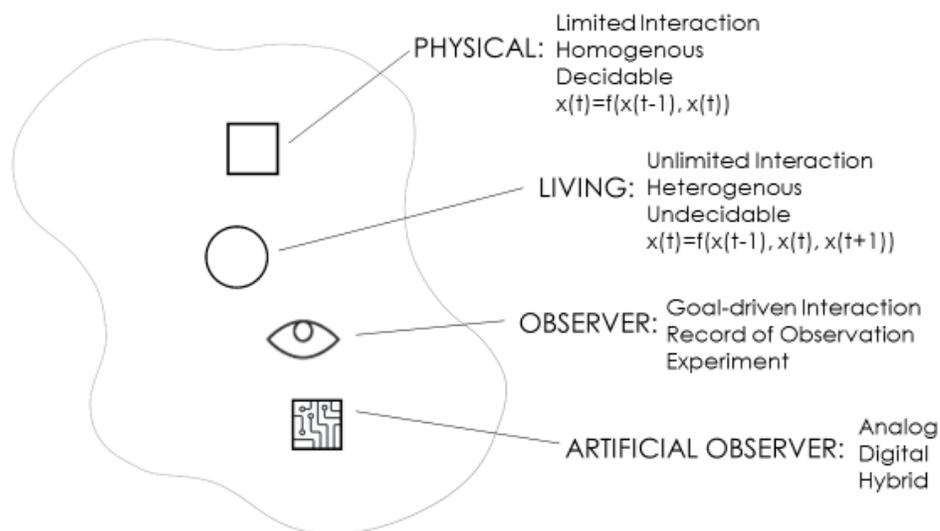

Fig. 3. Characterizing the physical and the living



The living, in its unlimited variety of ever-changing forms is G-complex, i.e., it is characterized by undecidability. For non-living physical entities, interaction takes the specific form of deterministic reaction, expressed in physical laws (such as those expressed in Newton's equations or in Einstein's theory of relativity), or embodied in measurement devices. Time is expressed as duration.

Change in the physical is unreflected: neither a stone nor a volcano is aware of the respective dynamics of how they change—and even less of why they change. They undergo change that can be observed, but which does not afford self-awareness. The observing subject, whose own condition is changed in the act of observation, takes note of their record of change (the duration sequence of the falling, the speed, the acceleration, the impact, or the detailed the sequence of a volcano's eruption). Based on empirical observation, the observer can design experiments pertinent to the dynamics of falling or of a volcanic eruption (or whatever other physical phenomenon). The living observes and, most important, can affect its own change. For the living, change is the outcome of interactions in which the physical (the dynamics of action-reaction) is complemented by anticipatory expression: current state contingent upon possible future state as it pertains to preserving life. Living entities (animals mostly) don't simply fall; they know (implicit knowledge) how to fall, as long as anticipatory processes take place. The example of falling extends to all organisms. Tardigrades (among other so-called cryptobiotic organisms, i.e., living in states of suspended dynamics), like seeds, also behave in an anticipatory manner. The process can be described (in an anthropomorphic way of speaking) as following the Möbius strip trajectory back to life [60]. Put otherwise, there is a continuum of life akin to the Klein bottle geometry (surface and interior are on the same plane), as there is in every creative endeavor. From inspiration to artwork—this is yet another example documented by the entire



history of art and literature. The possible future state in the dynamics of cell interaction or in neuronal synapses is of a different nature, and easier to understand than that of the tick that can be in "suspended" life (up to 18 years) until the butyric acid of a body in its *Umwelt* [61] triggers its action. Change in the living is reflected in the form of its representation through successive states. Living interaction is not reducible to the physical action-reaction sequence.

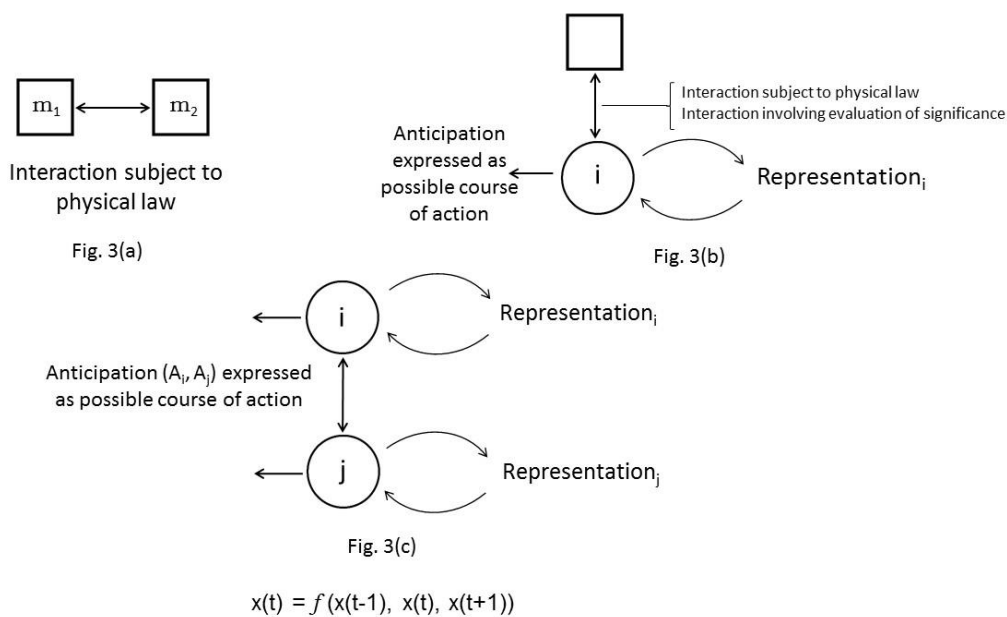

Fig. 4. Interaction and dynamics

The description of physical interaction conjures quantity: its observation (measurement) results in data. The description of living interaction conjures quality: its observation results in information, i.e., data associated with meaning [62]. Information, characteristic of life, is not physical [63]. Lived time and clocked time are different. In the physical, the scale of time is constant; in the living, it is variable.

Conclusion: *Change is the outcome of interaction*



Physics continuously provides experimental reproducible evidence concerning the cause-and-effect perspective expressed in determinism. If the physical world should start over (using Gould's suggestive replay of the film), its dynamics would, for all practical purposes, be the same. Evolution provides empirical evidence of its underlying process: anticipation. Successful anticipation drives survival. If life were to begin again (replay the video), the outcome would be different in each such re-beginning, and not predictable. The reactive and the anticipatory are integrated in the dynamics of life. Anticipation, as the underlying process of evolution, is documented through empirical evidence. It expresses characteristics of the living such as adaptivity, holism, purposefulness, and creativity, and it provides the premise for understanding emotion, the relation between the embodied brain and mind interaction, the focus on meaning.

To expect experiments involving the living (of interest not only to psychology, but also to the biomedical sciences and many other fields of inquiry pertinent to life) to be reproducible is epistemologically equivalent to reducing the living to its physical substratum, and biology to physics and chemistry (the reductionist doctrine). A particular form of this reduction is the machine model. Humans made the machine-god to replace the omnipotent, and after that they took the machine functioning predicated by the machine-god and committed to the practice of the machine-religion based on this belief. Experiments within this religion of the machine are reproducible, since they confirm the premise: the machine-god means that everything is a machine, or behaves like one, hence it is fully predictable.

Another aspect of the same scientific theology is that of conditioning: we created god, defined the commandments, and expect everyone to submit to them as though they come from a higher authority. It is conditioning, with a reward mechanism attached to it: respect the "commandments" as though they were "god-sent." The "god you made up" will reward you by



making you more god-like. In the machine-god phase, the same is practiced: the machine rewards those who accept that they are machines—and who behave accordingly—by making them more machine-like. Many experiments turn out to be mere instances of conditioning (psychology outperforms every other known discipline in this respect). Debunked many times over, conditioning became a feature of the very large body of "press the button" experiments—act like a  machine—meant to legitimize user interfaces, cognitive aspects of perception, efficiency of behavioral treatment.

In general, a limited understanding of causality as it pertains to life, i.e., transcends that of physics, dominates experiments concerning the living. By contrast, in modern physics, and not only in the quantum mechanics perspective, causality is gradually approached within a broader understanding of determinism and even openness to non-deterministic processes. Awareness of non-linearity and stochastic aspects of physical phenomena permeates such a view.

Mapping from an open system (extending from the cells to the whole human being), of extreme dynamics, to the closed system of the experiment--which by definition is supposed to be decidable—might result in reproducibility. But what is reproduced is a false assumption, not knowledge-bearing hypotheses about change. The validity of some 40,000 fMRI studies, and more broadly the interpretation of neuroimaging results, was recently questioned [64], after the fMRI (25 years old) technology itself was critically assessed ([65] among others). False-positive rates of up to 70% concerning its most common statistical methods, which have not been validated using real data, are actually a proof of a replicated misguided assumption.

But even within the mechanistic view of the living replication is by no mean guaranteed. As impressive as the Human Genome project was, it is a good example of irreproducible experiments (not to mention of its incompleteness). It was generated under the reductionist



assumptions of a blueprint—published as such [66]—of a *homo sapiens* that does not change over time, i.e., epigenetics ignored. What was extracted is a truncated image of gene syntax. The 1000 Genomes Project (2008-2015), aimed at studying variation (initially ignored) and genotype data, is an example of an improved understanding but yet still another irreproducible experiment. It affords useful empirical data, i.e., access to some semantic aspects of gene expression. The goal, probably not yet on the radar of scientific inquiry, should be the pragmatic level, where meaning is constituted in the context of life unfolding in an anticipatory manner. However, that would entail the need to accept that experiments within the decidable are reproducible. The complementary is evident: experiments concerning the dynamics of the G-complexity domain (the living) cannot be replicated. They are of extreme importance to science, but more as a source of data—the time (actually duration) series of the observed phenomenon. The synchronic view (bearing the time stamp of the experiment clock) begs for its complementary diachronic representation of integrated processes—involving a variety of clocks and timescales.

This idea is relatively well illustrated by the entire cycle of reproduction. Pregnancy [67] is a convincing example of anticipatory expression underlying creation, i.e., the birth of some entity that never existed before. For instance, anticipatory adaptations that diminish or eliminate the influence on the fetus of the mother's stressful experiences take quite a number of forms (from the flow of energy favoring the new life to triggering lactation). Oxytocin is released in advance of parturition and during lactation. Maternal behavior is also changed in anticipation of the birth proper. An altered emotional condition parallels new endocrine and cognitive functions.   As was already pointed out, the living, pregnant or not, is in a continuous state of remaking itself, *sui generis* re-creation of its constitutive cells—each different from the other—and thus of the entire organism. The constancy of physical (non-living) entities, even those of extreme dynamics (such



as black holes), stands in contrast to the variability of any and all organisms and the matter in which they are embodied.

An assumption similar to that of the Human Genome governs the current Connectome project. It will be ten or one hundred times more costly than the Genome project, but not more adequate in reporting on the variability of the cortex. Brain activity has become the showcase of computational modeling. There is, of course, much to gain from computational models in physics applications—the Juno space mission is only a recent spectacular example. In the biological realm an intrinsic limitation is ignored: algorithmic computation captures only the deterministic aspects of change. In addition, the premise that such processes are of algorithmic nature, was never proven—or at least questioned. The algorithmic is decidable; moreover, it is tractable. This means that the execution of the program representing the dynamics of the process represented by the computation takes a time represented by the polynomial of the steps required (for instance, for n steps, the polynomial function of n).

**Thesis 4:** The decidable can be represented by a tractable algorithmic computation.

**Thesis 5:** The living is not algorithmic.

**Thesis 6:** The undecidable is not tractable, neither in computational form nor in any form of data processing.

Even considering infinite computing resources of any kind of computation (analog, algorithmic, interactive, etc.)—which would undermine our current understanding of the relation between energy and data processing—the undecidable would remain intractable.



The ants (phildris nagasan) that "cultivate coffee for accommodation" accumulated experience—in this particular form of agriculture—over millions of years [68], way longer than the human being. An inverse computation (from today to when this ants' "agriculture" actually emerged), might explain the anticipatory characteristics of the elaborate process. But to perform it would take more than the energy involved in the evolution of life over that time. The physics of what the ants are doing is relatively simple; even an abacus would suffice for calculating how the process takes place.

The guaranteed reproducibility of computational neuroscience experiments conjures knowledge and validation not about the brain, whose deterministic and non-deterministic aspects complement each other in its functioning, but about algorithmic computation. Interactive computation, as well as other forms of computation, in line with the dynamics of interaction of the living in general, and of the brain in particular, are rarely considered [69, 70].

Windelband's [71] view of *nomothetic* science (expressed in universally valid laws, such as Newton's laws of mechanics) and *idiographic* science (diachronic processes subject to empirical observations) could as well guide in defining new methods for gaining knowledge peculiar to the living. Let us recall all those biologists (not only Gould, mentioned above) who have been questioning the assumption that there are laws that describe the living. Biologists mostly are experiencing the uniqueness of each subject and wonder how this uniqueness (idiographic characteristic) can be described. There is a direct practical consequence to this distinction: medical care as a reactive praxis of fixing the "human-machine," or individualized care (the art and science of healing) reflecting the awareness of uniqueness. A knee replaced is an example of an experiment replicated many times. A genetic-based treatment, of extreme individual nature, is as unique as the new attempts at immunotherapy in addressing conditions for which reactive



medicine is not an option. Moreover, the crisis of the experiment is also the crisis of our understanding of how knowledge is acquired, validated, and shared. Although the languages of visualization, modeling, and simulation are different from those of analytic expression, logic, mathematics, etc., we continue to expect some uniformity, as though the reality we question is by necessity homogenous. Peirce [72] suggested diagrammatic thinking as an alternative. So far, his ideas remain outside the mainstream of science. In recent years, Leamer [73], among others, contrasted "theory and evidence" vs. "patterns and stories." For some reason, neither biological theorists (such as Elsasser, Rosen, Pattee), and more recently Kaufmann [55], nor philosophers of the subject (in particular the competent Arran Gare [74]) has taken note of these developments. Even when the notion of *story* is mentioned, work in defining it (in contrast to the narrative) is ignored [75].

Consequently, *story* remains a rather suspicious candidate, although the uniqueness of life phenomena speaks more in favor of variety (which stories can offer) than the replication of experiments. The proponents of physics as "the science of everything," are grounded in its constructs. Those who advance alternative understandings of life processes know more about what they reject than what defines the culture of the subject of biology. For instance, they ignore contributions coming from outside their own context (such as those of researchers who worked in what used to be the Soviet Union, and who were severely censored). Just for the sake of the argument (i.e., integrating ideas from outside the culture shaped by the Cartesian view), let us mention that Bernstein [76] wrote about the "repetition without repetition" characteristic of the living as an expression of its dynamic variability. Machines provide mechanical repetition, which is their expected performance. On account of empirical evidence, this is yet another argument in favor of finally transcending the machine view characteristic of Cartesian determinism and



reductionism. Gelfand's [77] take on the matter continuing Wigner's line (on the effectiveness of mathematics in physics) points in the same direction: "There is only one thing which is more unreasonable than the unreasonable effectiveness of mathematics in physics, and this is the unreasonable ineffectiveness of mathematics in biology." Mathematics captures the decidable. Other descriptions, such as the record of a process, return testimony to the undecidable. Progress in science renders the need for a "new Cartesian revolution," at the forefront of science's efforts to better understand change in the specific manner in which it characterizes life.

**There are no conflicting interests to be reported.**

## ACKNOWLEDGMENTS

This study is the outcome of a long term endeavor. Interactions with distinguished colleagues and many young researchers helped in defining the foundation for this work. The author would like to acknowledge Robert Rosen for his pioneering work in defining the living, and colleagues from the University of California–Berkeley, Professors Harry Rubin (Biology) and Lotfi Zadeh (Electrical Engineering and Computer Science); Professor Solomon Marcus (mathematician, Member of the Romanian Academy), Aloisius H. Louie, Stuart Kauffman, Kalevi Küll (University of Tartu, Estonia), and more recently Arran Gare (*Swinburne* University of Technology, Melbourne, Australia). for their intellectual openness to new ideas and encouragement.